# Transforming agrifood production systems and supply chains with digital twins

*Asaf Tzachor* [1,2], *Catherine E. Richards*[1,3]*, Scott Jeen*[3,4]

## Abstract

Digital twins can transform agricultural production systems and supply chains, curbing greenhouse gas emissions, food waste and malnutrition. However, the potential of these advanced virtualization technologies is yet to be realized. Here, we consider the promise of digital twins across five typical agrifood supply chain steps and emphasize key implementation barriers.

## Main

Agrifood production systems and supply chains are currently not on track to meet the sustainable development goals. They are wasteful and polluting, breach several of the so-called planetary boundaries, and fail on their most basic premise to provide an expanding global population with safe and nutritious diets, leaving some 900 million people undernourished.[1]

As a response, transformation through digital technological innovation is often proposed.[2,3,4,5] In such proposals, computer-enabled technologies, including smart sensors, artificial intelligence (AI) and other embedded systems, are fundamental. Here, we discuss the potential of digital twin (DT) technology, which despite its potency and increasing diffusion across industrial domains has not been considered for the purpose of improving agrifood sector sustainability, namely through mitigating malnutrition and undernutrition, reducing greenhouse gas (GHG) emissions and preventing food waste.

### Advantages of virtualized agrifood systems and supply chains

DTs are virtual representations of living or non-living physical entities. Enabled by improvements in computing capabilities, they exist *in silico*, that is as computer simulated models.[6] Deployment of sensors that detect biological, chemical, and physical properties of objects in real-time, ensures that the digital counterparts of these measured objects are accurate and 'live'.[7] In such cyber-physical architectures, changes that occur in the physical system are modifying its virtual twin simultaneously and continuously.

With origin in experimental designs of satellites, spacecrafts, city infrastructures,[8] and civil engineering writ large, in recent years DTs have been re-purposed to address predicaments such as climate change and extreme weather.[9]

By simulating the state of physical systems, DTs can be queried using advanced modelling techniques to uncover optimal behavior. Reinforcement Learning (RL), a subfield of AI that enables autonomous agents to make decisions in complex systems[10], can be deployed in DTs to advise optimal control strategies to the physical counterpart. RL agents take the current state of a system as input, and predict future action sequences that optimize system behavior. DTs allow agents to simulate many control sequences to determine which aligns best with the control objective before advising the physical system.

Combining virtual replicas with such advanced decision-making technologies will have profound transformative implications for the agrifood sector, offering possible remedies to the problems of malnutrition, GHG emissions, and food waste. To appreciate these prospects, we acknowledge potential applications across five supply chain steps: (a) agricultural inputs, (b) primary agricultural production, (c) storage and transportation, (d) food processing, and (e) distribution and retail.

*Inputs for agricultural production*

Agricultural inputs commonly refer to agro-chemicals, such as nitrogen (N) and phosphorous (P) fertilizers, pesticides, and crop seeds, which are essential for yield productivity. The carbon footprint involved in manufacture of these inputs is considerable. For example, $CO_2$ emissions of N fertilizer production in China is estimated at 452 Tg $CO_2$-eq, constituting 7% of total GHG emissions from the Chinese economy. Measures to improve heat conversion efficiency in power plants supporting N fertilizer manufacturing are recognized as an essential intervention to lower carbon intensity.[11] In this context, 'virtual power plants' could be developed and used by RL agents to find control policies that maximize electricity generation whilst minimizing $CO_2$ emissions.[12]

DTs proven to operate at the molecular, cell, tissue and organ levels[13] can enable precise simulations of crops. New 'virtual crops' could be rapidly stress-tested in computer laboratories under alternate conditions, including precipitation, temperature and salinity, to discover desirable traits and develop genetically engineered seeds of climate-resilient staples.

*Primary agricultural production*

Beyond the organ level, virtualization of entire farming systems that replicate atmospheric factors, geomorphological processes and edaphic conditions, including soil microbiology, would support precision agriculture at unprecedented scales. Such DTs are likely to use cameras and sensors to sample humidity, moisture content, temperature, irradiance, irrigation and nutrient supply as often as every minute. RL agents could use these measurements to form a state representation and recommend irrigation, lighting and nutrient control actions toward some objective, likely minimization of resource-use whilst maximizing crop yield.[14] For example, DTs could suggest optimal application of pesticides, currently amounting to 3.5 million tonnes utilized annually worldwide, oftentimes excessively.

Moreover, DTs may promote rewilding, sediment trapping and additional nature-based solutions for land management and restoration,[15] through rapid experimentation in 'virtual farms'. *In silico* 'what-if' simulations could elicit further benefits, such as testing and identifying pathways to increase carbon sequestration in croplands and pastures, or using agro-forestry techniques, such as integrated green belts for wildfire prevention.

As in other domains, including water and electricity infrastructure, DTs can support predictive maintenance,[16] for instance, of irrigation systems in plantations to minimize food losses. In intensive closed-environment agriculture (CEA), such as commercial aeroponic greenhouses and hydroponic systems, DTs may be used in structure design and operations to suggest optimum light intensity, humidity, temperatures, $CO_2$ concentrations and water-nutrient recycling.

*Storage and transportation*

Commodity chains that connect local produce to markets typically involve transit in freight trains and bulk carriers as well as temporary storage in terminal elevators. In rail, road and sea vessels, and in storage silos, cargos of grain are susceptible to mold, mustiness and early germination.

Ventilation management is essential to prevent dampness and fungal infestation, such as *Aspergillus* and *Penicillium* that frequently deteriorate the quality of cereal bulks.[17] DTs already employed for improved HVAC systems design[18] could be re-purposed to this end. Additionally, real-time replicas of stationary elevators and vessels on voyage could track ventilation periods and moisture content of cargo as well as provide early warning of mycotoxin contamination that warrants fumigation.

In cold chains of perishable produce, where fruit, vegetable, dairy, meat and seafood products are pre-cooled and provisionally stored in refrigerated facilities, computer simulations may advise on energy efficiency measures to reduce carbon emissions. Synchronized DTs can monitor food temperatures, humidity, delivery schedules, respiratory behavior, and grid carbon intensity; RL agents can then optimize the control of cooling equipment to draw power from the grid when carbon intensity is lowest to minimize emissions whilst maintaining food quality[19].

*Food processing*

Paired with sensing technologies, DTs can be integrated across food processing and packaging facilities that convert agricultural commodities, such as corn or cattle, to ingredients and end-user food products, including tinned vegetables, meat cuts, ready meals and confectionery.

Food loss and waste in this echelon are prevalent in both developed and developing regions, with implications for food security and the environment.[20] In the UK, for example, food waste in this echelon stands at five megatonnes each year.[21]

Here, DTs can support industrial ecology approaches to prevent food loss, in the same way they have been used to enhance circular economy applications in construction manufacturing.[22] Such DTs could monitor ingredient delivery schedules, plant throughput, ingredient wastage, operator work schedules and demand forecasts. RL models could be trained to manipulate manufacturing equipment to match food processing to expected demand whilst minimizing waste.

*Distribution and retail*

Food distribution networks are significant contributors to global GHG emissions, with food retail alone responsible for approximately 0.3 gigatonnes of $CO_2$ annually.[23] Food discarded in this echelon is considerable too, for example, with estimates suggesting 366 kilotonnes of food waste per year in the UK.[24] These losses are attributed to inefficient warehousing, hypermarkets and supermarkets operations including shelf management and failure to monitor and measure food waste.[25]

DTs mimicking food distribution systems could be used to optimize delivery schedules to minimize carbon emissions and food wastage. Such DTs could monitor the location of delivery vehicles across the road network, food inventory in retail stores, food embodied emissions traffic, weather and shelf-life of food in transit.

Given this state representation, an RL agent may decide that minimizing food wastage, and thus system-level emissions, could be achieved by sending food to a retailer further from the distribution centre with low inventory levels, rather than a closer store more likely to incur wastage. Recent reviews suggest these simulations could further predict delays in supply chains, signs of food spoilage and potential food losses as well as recommend preventative measures.[26]

Where discard of food surplus is expected, the expansion of DTs to encompass networks of food re-distribution, such as community soup kitchens, can aid in waste mitigation and improving the nutritional security of vulnerable populations. Such expansion may also include growers to more effectively apportion and dispense unharvested produce.

# Enabling and disabling factors for virtualized agrifood value chains

'Live' DTs offer comprehensive computational ecosystems for simulating crops, farms, agricultural equipment, storage facilities, processing factories, and distribution networks. Nevertheless, agrifood stakeholders must be cognizant of techno-economic limitations if deployment of DTs is to be realized successfully and at scale.

First, robust virtual replicas rely on two elements: (a) appropriate sensor coverage and (b) model uncertainty quantification. For advanced decision-making systems to recommend optimal control strategies using a DT, its sensors must be sufficiently predictive of the agent's objectives. For example, a DT of an agrifood storage facility could only be used to predict food spoilage if it monitors correlating variables, like temperature, food type and product age. Even with sufficient sensor coverage, the DT can only ever be an approximation of the physical system meaning its state representation and future predictions are uncertain. In response, several authors recommend building DTs using Bayesian methods, but robust methods for dealing with DT uncertainty and decision making remains an open challenge.[27] Deploying DTs that capture uncertainty explicitly is crucial to mitigating these issues.

In the same vein, setting 'live' replicas of entire supply chains that encompass re-distribution centers, such as food banks and soup kitchens in lower-income communities, would require hefty investments in data architectures, including cloud computing and on-premise sensors.

However, it is likely that private firms at the forefront of DTs research and development would lack incentive to invest in cyber-physical systems that promote ecological and humanitarian causes, such as agro-biodiversity and food rescue, but yield no direct financial returns. This may stifle the dissemination of DTs for agrifood sector transformation, particularly in areas where digital innovation is needed the most.

Second, modelling flaws may be introduced in design, through human error in coding or merging error-free but discordant algorithms or data. A small notational error in the code of a computational model used for predictive maintenance of an irrigation system, for instance, could result in ill-informed decisions leading to crop yield failures and produce loss.

Third, lack of common modelling standards for DTs might create compatibility difficulties in integrating separately created models.[28] For example, patching a virtual representation of a new piece of cooling equipment in cold chains, programmed by the manufacturer to monitor temperature in degrees Fahrenheit, into an existing cold chain that regulates temperature in degrees Celsius will result in immediate food spoilage.

*Lifting barriers*

A concentrated effort by international and public institutions is necessary to guarantee that DTs are implemented outside of their origin context in civil and mechanical engineering to fulfil their promise in agrifood sector transformation.

Nonprofit international research centers, such as CGIAR and its Platform for Big Data in Agriculture, ought to be financed to promote open-access and standardized datasets that could support DTs from molecular to landscape levels, including of orphan crops and indigenous agro-ecologies as well as to develop open-source and secured platforms for agricultural DTs initiatives. Public institutions should further invest in underlying standards and data architectures along value chain echelons, deploy bio-physical and bio-chemical smart sensors, telecommunication networks, and cloud computing to meet the data storage and processing demands of DTs.

At the same time, it is essential to expand computer science training and skills with designated syllabi and simulation software for actors involved in the agrifood sector, including in technical and vocational education and training (TVET) programs. This should facilitate a DT spillover across disciplines, domains and geographies.

Investments should prioritize lower-middle income economies, where the greatest number of smallholders operate, rural credit markets are immature, agricultural productivity is low, food spoilage and waste are widespread, and malnutrition is prevalent. Without deliberate institutional effort, underdeveloped regions will likely be the slowest to benefit from DTs, much in the same way Green Revolution technologies have bypassed the most vulnerable.[29]

Finally, DTs that already inform scientists and engineers in other domains should be adapted promptly and with robust governance to achieve agrifood production system and supply chain sustainability.

# Acknowledgements

This paper was made possible through the support of a grant from Templeton World Charity Foundation, Inc. The opinions expressed in this publication are those of the authors and do not necessarily reflect the views of Templeton World Charity Foundation, Inc.

# Author contribution

A.T., C.E.R., and S.J. developed the paper jointly, and contributed equally to the writing of the text.

# Author information

Affiliations

[1] Centre for the Study of Existential Risk (CSER), University of Cambridge, Cambridge, UK.

[2] School of Sustainability, Reichman University, Herzliya, Israel.

[3] Department of Engineering, University of Cambridge, Cambridge, UK.

[4] Alan Turning Institute, London, UK.

# Competing interests

Authors declare no competing interests.